\documentclass[12pt]{article}
\usepackage{epsfig}
\usepackage{amsfonts,amssymb}
\usepackage{latexsym}
\usepackage[OT2,OT1]{fontenc}
\usepackage[ansinew]{inputenc}

\def\cyr{\fontencoding{OT2}\fontfamily{wncyr}\selectfont}

\def\G{\Gamma}
\begin{document}

\title{Ambiguity-free formulation of the Higgs-Kibble model.}
\maketitle
\large \rm
\begin{center}
\vskip 0.7 truecm A.~Quadri$^{a,b}$\footnote{e-mail: {\tt
andrea.quadri@mi.infn.it}}, A.~A.~Slavnov$^{c}$\footnote{e-mail:
{\tt slavnov@mi.ras.ru}}
\end{center}

\normalsize
\begin{center}
$^a$
Physikalisches Institut\\
Albert-Ludwigs Universit\"at Freiburg\\
Hermann-Herder-Strasse 3a,\\
D-79104 Freiburg i.Br., Germany\\
$^b$
Dip. di Fisica, Universit\`a degli Studi di Milano\\
via Celoria 16, I-20133 Milano, Italy\\
$^c$
Steklov Mathematical Institute\\
Gubkina st.8, Moscow, Russia
\end{center} \maketitle
\vskip 0.8 truecm

\begin{abstract}
\noindent
A renormalizable ambiguity-free formulation of the Higgs-Kibble
model is proposed.
\end{abstract}

\vskip 5 truecm
\leftline{FR-PHENO-2010-029}
\newpage
\section {Introduction}

The problem of ambiguity in the choice of a gauge condition in
nonabelian gauge theories is usually associated with the massless
Yang-Mills field \cite{VG}, \cite{IS}. In this case
%%%%
%%%%
the problem of ambiguity is somewhat academic, as the
scattering matrix acting in the space of color asymptotic states does not
exist because of infrared singularities and the notion of unitarity in
the asymptotic space makes no sense.

However the question about an ambiguity in the choice of a gauge
condition arises also in the Higgs-Kibble model \cite{H}, \cite{Ki},
where infrared singularities are absent and the scattering matrix is
well defined in asymptotic space. But the Higgs model in
renormalizable gauges like $\partial_{\mu}A_{\mu}=0$ suffers from
the same ambiguity problems as the pure Yang-Mills theory. In the
unitary gauge the ambiguity may be easily removed by a redefinition
of the fields, but the theory in this
%%%%
gauge
%%%%%
is
not renormalizable.

Recently in the papers (\cite{Sl1}, \cite{Sl2}, \cite {QS}) the new
formulation of the Yang-Mills theory was proposed, which allows a
unique fixation of the gauge. Therefore the quantization in this
gauge makes sense both in perturbation theory and beyond it.
Moreover in the paper (\cite{QS}) it  was shown that the
perturbation theory in the ambiguity-free gauge is renormalizable,
however the renormalization is not reduced to a multiplicative
redefinition of the parameters of the effective action but includes
nonmultiplicative redefinition of the fields.

In this paper a similar procedure will be constructed for the
Higgs-Kibble model. We shall show that this model may be formulated
in a fashion, analogous to the one proposed in the papers
(\cite{Sl1}, \cite{Sl2}, \cite {QS})for the pure Yang-Mills field. A
possibility of a unique choice of the gauge arises, and the
perturbation theory in this gauge is explicitly renormalizable.

\section {The Higgs-Kibble model in the ambiguity free gauge}

We start with the explicitly gauge invariant model, described by the
Lagrangian
\begin{eqnarray}
L=-
\frac{1}{4}F_{\mu\nu}^aF_{\mu\nu}^a+(D_{\mu}\varphi^+)^*(D_{\mu}\varphi^-) + (D_{\mu}\varphi^-)^*
(D_{\mu}\varphi^+)\nonumber\\
+(D_{\mu}\varphi)^*(D_{\mu}\varphi)-\lambda^2(\varphi^* \varphi-\mu^2)^2\nonumber\\
-[(D_{\mu}b)^*(D_{\mu}e)+(D_{\mu}e)^*(D_{\mu}b)] \, . \label{1}
\end{eqnarray}

In the present paper we refer to the gauge group SU(2),
however the construction can be generalized to
other groups.
Here the field $\varphi$ is a complex doublet describing the Higgs meson,
and the fields $\varphi^+, \varphi^- $ are auxiliary fields which form
analogous doublets, conveniently parametrized by the Hermitean components
\begin{equation}
 \Phi=\left( \frac{i\Phi_1+\Phi_2}{\sqrt{2}}, \frac{\Phi_0-i\Phi_3}{\sqrt{2}}\right)
\label{2}
\end{equation}
The scalar fields $b$ and $e$ may be also described by the complex
doublets, however their components are anticommuting and to provide
Hermiticity of the Lagrangian (\ref{1}) it is necessary to take the
components of the field $b$ antihermitean. $D_{\mu}$ denotes the usual
covariant derivative.

It is easy to see that in the vacuum sector for the fields
$\varphi^{\pm},b,e$ the path integral for the scattering matrix
corresponding to the Lagrangian (\ref{1}) reduces to the path
integral for the usual Higgs-Kibble model. In the integral
\begin{equation}
\int \exp\{i\int L ~ d^4x\}d\mu \label{3} \end{equation} where
$d\mu$ includes also the product of the differentials of the
auxiliary fields \textbf{$\varphi_{\pm},b,e$}, one may integrate
explicitly over these fields. After that the integral (\ref{3})
coincides with the path integral for the Higgs-Kibble model.

We however are going to do in the Lagrangian (\ref{1}) a shift of the
fields $\varphi$ and $\varphi^-$, leading to a spontaneous breaking of
the symmetry. Such a transformation is not an admissible change of
variables in the integral (\ref{3}), as it changes the asymptotic values
of the integration variables. Therefore the unitarity of the "shifted"
theory requires a special study.

After the shift
\begin{equation}
\varphi^-(x) \rightarrow \varphi^-(x)- \hat{m}; \quad
\varphi(x)\rightarrow \varphi(x)- \hat{\mu}
\label{4}
\end{equation}
where $\hat{m}$ and $\hat{\mu}$ are the constant spinors
\begin{equation}
\hat{m}=(0, m/g); \quad \hat{\mu}=(0, \mu/g)
 \label{5}
\end{equation}
the Lagrangian (\ref{1}) acquires the form
\begin{eqnarray}
L=-
\frac{1}{4}F_{\mu\nu}^aF_{\mu\nu}^a+(D_{\mu}\varphi^+)^*(D_{\mu}\varphi^-)+
(D_{\mu}\varphi^-)^*(D_{\mu}\varphi^+)\nonumber\\
-[(D_{\mu}\varphi^+)^*(D_{\mu}\hat{m})+(D_{\mu}\hat{m})^*D_{\mu}\varphi^+]\nonumber\\
-[(D_{\mu}b)^*(D_{\mu}e)+(D_{\mu}e)^*(D_{\mu}b)]+(D_{\mu}\varphi)^*(D_{\mu}\varphi)\nonumber\\
-[(D_{\mu}\varphi)^*(D_{\mu}\hat{\mu})+(D_{\mu}\hat{\mu})^*(D_{\mu}\varphi)]\nonumber\\+(D_{\mu}
\hat{\mu})^*(D_{\mu} \hat{\mu})-\lambda^2[(\varphi- \hat{\mu})^*(\varphi-
\hat{\mu})-\mu^2]^2 \, . \label{6}
\end{eqnarray}
The last terms in this equation, starting from the term
$(D_{\mu}\varphi)^*(D_{\mu}\varphi)$ coincide identically with the
corresponding terms of the Higgs-Kibble model. In  particular the
presence of the term $(D_{\mu} \hat{\mu})^*(D_{\mu} \hat{\mu})$
results in the mass term for the Yang-Mills field
$\frac{\mu^2}{2}A_{\mu}^aA_{\mu}^a$.

The Lagrangian (\ref{6}) being obtained by the shift from the gauge
invariant lagrangian (\ref{1}) obviously is invariant with respect to the
"shifted" gauge transformations. In particular the fields $\varphi^a_-$
and  $\varphi^a$ are transformed as follows
\begin{eqnarray}
\delta
\varphi_-^a=m\eta^a+\frac{g}{2} \epsilon^{abc}\varphi_-^b \eta^c+
\frac{g}{2} \varphi_-^0\eta^a\nonumber\\
\delta \varphi^a=\mu\eta^a+\frac{g}{2} \epsilon^{abc}\varphi^b \eta^c+
\frac{g}{2} \varphi^0\eta^a \label{7}
\end{eqnarray}
As the fields $\varphi^a$ and $\varphi^a_-$ under the gauge
transformations are shifted by arbitrary functions $m \eta^a$ and $\mu
\eta^a$ one can impose a gauge condition on any of these fields.

If one makes firstly in the classical Lagrangian the following change of
variables
\begin{eqnarray}
\varphi^0_-=\frac{2m}{g}(\exp\{\frac{gh}{2m}\}-1); \quad
\varphi^a_-=\tilde{M}\tilde{\varphi}^a_-\nonumber\\
\varphi^a_+=\tilde{M}^{-1}\tilde{\varphi}^a_+; \quad \varphi^0_+=
\tilde{M}^{-1}\tilde{\varphi}^0_+\nonumber\\
e= \tilde{M}^{-1} \tilde{e}; \quad b= \tilde{M} \tilde{b} \label{8}
\end{eqnarray}
 where
\begin{equation}
\tilde{M}=1+ \frac{g}{2m}\varphi^0_-  = \exp\{\frac{gh}{2m}\} \label{9}
\end{equation}
and impose the gauge condition
\begin{equation}
\tilde{\varphi}^a_-=0 \label{10}
\end{equation}
then under the gauge transformations
\begin{equation}
\delta \tilde{\varphi}^a_-=m \eta^a \label{11}
\end{equation}
and Gribov ambiguity is absent.

The effective Lagrangian in the gauge (\ref{10}) looks as follows
\begin{eqnarray}
\tilde{L}=-\frac{1}{4}F^a_{\mu \nu}F^a_{\mu \nu}+
\partial_{\mu}h \partial_{\mu} \tilde{\varphi}^0_+-
\frac{g}{2m}\partial_{\mu}h \partial_{\mu}h \tilde{\varphi}^0_+\nonumber\\
+m \tilde{\varphi}^a_+ \partial_{\mu}A_{\mu}^a-[((D_{\mu} \tilde{b})^*+
\frac{g}{2m} \tilde{b}^*\partial_{\mu}h)(D_{\mu} \tilde{e}-
\frac{g}{2m} \tilde{e}\partial_{\mu}h)+h.c.]\nonumber\\
+ \frac{mg}{2}A_{\mu}^2 \tilde{\varphi}^0_++ g \partial_{\mu}h A_{\mu}^a
\tilde{\varphi}^a_++ \frac{\mu^2}{2}A_{\mu}^aA_{\mu}^a+\mu \varphi^a
\partial_{\mu}A_{\mu}^a+\ldots \label{12}
\end{eqnarray}

Out of the terms which arise in the Higgs-Kibble model
after the shift of the field $\varphi^0$ we have only
displayed explicitly in eq.(\ref{12}) the mass term
for the gauge field and the Goldstone-gauge boson
bilinear. The other terms are denoted by $\ldots$.

As
one sees the free Lagrangian corresponding to (\ref{12}) differs
from the free Lagrangian corresponding to the massless Yang-Mills
field by the presence of the mass term
$\frac{\mu^2}{2}A_{\mu}^aA_{\mu}^a$, the mixed term
$\mu\varphi^a\partial_{\mu}A^a_{\mu}$ and the free Lagrangian for
the Higgs field $\varphi$. The mixed term may be removed by the
change of variables
\begin{equation}
\tilde{\varphi}_+^a\rightarrow  \tilde{\varphi}_+^a-
\frac{\mu}{m}\varphi^a\label{13}
\end{equation}

Resulting free propagators look as follows
\begin{eqnarray}
\Delta(A^a_{\mu}A^b_{\nu})=\frac{-i\delta^{ab}}{p^2-\mu^2}T_{\mu\nu},
\quad \Delta(A^a_{\mu}\tilde{\varphi}^b_+)=-\delta^{ab}
\frac{p_{\mu}}{mp^2},\quad
\Delta(\tilde{\varphi}^0_+h)=\frac{i}{p^2}\nonumber\\\Delta(\tilde{b}^0
\tilde{e}^0)=\frac{i}{p^2}, \quad \Delta(\tilde{b}^a
\tilde{e}^b)=\frac{i\delta^{ab}}{p^2}, \quad \Delta(\varphi^0
\varphi^0)=\frac{i}{p^2-M_H^2}, \quad \Delta(\varphi^a \varphi^b)=
\delta^{ab}\frac{i}{p^2} \label{14}
\end{eqnarray}
Here $T_{\mu\nu}$ is the transversal projector. We also set
$M_H^2=4\lambda^2 \mu^2$ for the mass squared of the physical Higgs
mode. It is not difficult to calculate the divergency index of an
arbitrary diagram. It is equal to
\begin{equation} n =
4-2L_{\varphi^0_+}-2L_{\varphi^a_+}-L_A-L_e-L_b-L_h-L_{\varphi^a}-L_{\varphi^0}
\label{15}
\end{equation}

The divergent diagrams may have no more than four external lines, that is
the model is manifestly renormalizable.

\section{Unitarity}\label{unit}

The Lagrangian obtained from eq.(\ref{6}) after the change (\ref{4})
includes a number of unphysical exitations corresponding to the ghost
fields $\varphi_+^a, \varphi_+^0, h, \varphi^a, e, b$ and zero components
of the Yang-Mills field. It is necessary to show that the scattering
matrix nevertheless is unitary in the physical sector which includes only
three components of the massive Yang-Mills field and one massive Higgs
scalar.

As in the case of the massless Yang-Mills theory the crucial role here is
played by the supersymmetry of the Lagrangian (\ref{1}). This Lagrangian
is invariant with respect to the supersymmetry transformations
\begin{eqnarray}
\delta \varphi^a_-=- b^a\nonumber\\
\delta \varphi^0_-=- b^0\nonumber\\
\delta e^a=\varphi^a_+\nonumber\\
\delta e^0=\varphi^0_+\nonumber\\
\delta b=0 \nonumber \\
\delta \varphi_+ = 0 \label{16}
\end{eqnarray}

 In terms of the
variables $ \tilde{\varphi}$ the supersymmetry transformations acquire
the form
\begin{eqnarray}
\delta h=- \tilde{b}^0 \nonumber\\
\delta \tilde{\varphi}^a_-=- \tilde{b}^a+ \frac{g}{2m}\tilde{\varphi}^a_-\tilde{b}^0\nonumber\\
\delta \tilde{\varphi}^a_+=- \frac{g}{2m} \tilde{b}^0 \tilde{\varphi}^a_+\nonumber\\
\delta \tilde{\varphi}^0_+=-\frac{g}{2m} \tilde{b}^0 \tilde{\varphi}^0_+\nonumber\\
\delta \tilde{e}^a=\tilde{\varphi}^a_++ \frac{g}{2m} \tilde{e}^a \tilde{b}^0\nonumber\\
\delta \tilde{e}^0=\tilde{\varphi}^0_++\frac{g}{2m} \tilde{e}^0 \tilde{b}^0\nonumber\\
\delta \tilde{b}^a= -\frac{g}{2m} \tilde{b}^a \tilde{b}^0\nonumber\\
\delta \tilde{b}^0=0 \label{17}
\end{eqnarray}

The Lagrangian (\ref{6}) after the change of variables (\ref{8}) is
invariant with respect to the corresponding BRST transformation and the
supersymmetry transformation (\ref{17}). Obviously it is also invariant
with respect to the simultaneous change of the fields combining these two
transformations. It allows to use instead of the canonical gauge fixing
the following gauge fixing
\begin{equation}
s^1 \int d^4x ~ \bar{c}^a \tilde{\varphi}^a_-= \int d^4x(\lambda^a
\tilde{\varphi}^a_-- \bar{c}^a(M^{ab}c^b- \tilde{b}^a) ) \, , \label{18}
\end{equation}
where $s_1$ is the nilpotent operator, similar to the BRST operator,
defined by the gauge transformation, leaving invariant the Lagrangian
(\ref{6})written in terms of the transformed variables and
\begin{eqnarray}
(s_1c)^a=-  \frac{g}{2}\varepsilon^{abc}c^bc^c\nonumber\\
(s_1 \bar{c})^a=  \lambda^a\nonumber\\
(s_1\lambda)^a=0 \, .\label{19}
\end{eqnarray}
Note that this modification does not change our conclusion about the
renormalizability of the theory.

 Therefore the scattering matrix is given
by the path integral
\begin{equation}
S= \int \exp\{i \int[L+\lambda^a
\tilde{\varphi}^a+m\bar{c}^ac^a+\tilde{b}^a \bar{c}^a]d^4x\}d\mu
 \label{20}
\end{equation}
where the boundary conditions are imposed on all the fields entering
the effective action except for the ultralocal ghost fields
$\bar{c},c,\lambda$. Performing explicit integration over
$\bar{c}^a, c^a$ we obtain in the exponent the effective action
which is invariant with respect to the modified BRST transformations
corresponding to the usual BRST transformations and the
supersymmetry transformations (\ref{17}) after the substitution
$c^a=\frac{\tilde{b}^a}{m}$:
\begin{eqnarray}
\delta A_{\mu}^a=\frac{1}{m}(D_{\mu}\tilde{b})^a\nonumber\\
\delta\tilde{\varphi}_-^a=0\nonumber\\
\delta h= -\tilde{b}^0\nonumber\\
\delta \tilde{\varphi}_+^a= \frac{g}{2m} \tilde{\varphi}^0_+ \tilde{b}^a+
\frac{g}{2m} \varepsilon^{abc} \tilde{\varphi}_+^b \tilde{b}^c-
\frac{g}{2m} \tilde{\varphi}^a_+\tilde{b}^0\nonumber\\
\delta\tilde{\varphi}_+^0=-\frac{g}{2m}( \tilde{\varphi}^a_+ \tilde{b}^a+
\tilde{\varphi}^0_+ \tilde{b}^0 ) \nonumber\\
\delta \tilde{e}^a= \frac{g}{2m}( \tilde{e}^a \tilde{b}^0 -
\tilde{e}^0 \tilde{b}^a- \varepsilon^{abc} \tilde{e}^b \tilde{b}^c)+
\tilde{\varphi}^a_+\nonumber\\
\delta \tilde{b}^0=0\nonumber\\
\delta \tilde{e}^0=\frac{g}{2m}( \tilde{e}^a \tilde{b}^a+ \tilde{e}^0
\tilde{b}^0)+ \tilde{\varphi}^0_+\nonumber\\
 \delta \tilde b^a=-
\frac{g}{2m}
\varepsilon^{abc} \tilde{b}^b \tilde{b}^c\nonumber\\
 \delta\varphi^a=\mu
\frac{\tilde{b}^a}{m}+\frac{g}{2m}\varepsilon^{abc}\varphi^b
\tilde{b}^a+\frac{g}{2m}\tilde{b}^a\varphi^0\nonumber\\
\delta\varphi^0=-\frac{g}{2m}\tilde{b}^a\varphi^a
 \label{21}
\end{eqnarray}
The invariance of the effective action with respect to the transformation
(\ref{21}) according to Noether theorem leads to the existence of the
conserved nilpotent charge $Q$, which allows to separate the physical
space by requiring its annihilation by the operator $Q$:
\begin{equation}
Q|\Phi>_{phys}=0 \label{22}
\end{equation}
where $|\Phi>_{phys}$ cannot be presented in the form
\begin{equation}
|\Phi>_{phys} = Q |\Psi> \, . \label{22.1}
\end{equation}

For asymptotic states this condition is reduced to
\begin{equation}
Q^0|\Phi>_{as}=0 
\, , \qquad |\Phi>_{as} \neq Q^0 |\Psi> \, ,
\label{23}
\end{equation}
where $Q^0$ is the free charge acting on the fields as follows
\begin{eqnarray}
Q^0A^a_{\mu}=\frac{\partial_{\mu} \tilde{b}^a}{m} \nonumber\\
Q^0 \tilde{e}^a= \tilde{\varphi}^a_+\nonumber\\
Q^0 \tilde{\varphi}^a_+ = 0 \nonumber\\
Q^0\varphi^a= \frac{\tilde{b}^a}{m}\nonumber\\
Q^0 \tilde{b}^a = 0 \nonumber\\
Q^0 \varphi^0 = 0 \nonumber \\
Q^0 \tilde{e}^0=\tilde{\varphi}^0_+\nonumber\\
Q^0 \tilde{\varphi}^0_+=0\nonumber\\
Q^0 h= -\tilde b^0\nonumber\\
Q^0 \tilde{b}^0=0 \label{24}
\end{eqnarray}
If one identifies the field $\tilde{e}^a$ with the antighost field
$\bar{c}^a$ in the ordinary Higgs-Kibble model, and the field
$ \frac{1}{m}\tilde{b}^a$ with the ghost field $c^a$, the first
 six transformations coincide with the BRST transformations in the
Higgs-Kibble model, thus providing the decoupling of the
fields $\tilde{\varphi}^a_+,\tilde{e}^a, \tilde{b}^a,\varphi^a$ and
unphysical components of the Yang-Mills field from the physical
states. The remaining transformations provide the decoupling of the
fields $e^0,h,b^0,\varphi^0_+$. Ultralocal fields $\lambda^a,
\bar{c}^a, c^a$ do not contribute to the asymptotic states.

Therefore our model has the same spectrum of observables as the
standard Higgs-Kibble model. However to conclude that our model is
equivalent to the usual one, one should prove that renormalization
preserves the formal relations obtained above.

\section{Renormalization}

The action $S=\int d^4x \, {\tilde L}$, where $\tilde{L}$ denotes
the effective Lagrangian corresponding to the effective action in
the exponent of (\ref{20}) after integration over $\bar{c},c$, is
invariant under the transformations eq.(\ref{21}) and under a global
SU(2) symmetry of the fields $A^a_\mu, \tilde \varphi_+^a, \tilde e,
\tilde b, \varphi$. However there are further terms that respect the
invariance under the transformations (\ref{21}) and the residual
global SU(2) invariance and are not forbidden by power
counting. The first one does not involve the Higgs doublet
$\varphi$ and is common to the massless Yang-Mills theory quantized
in the ambiguity-free gauge \cite{QS}
\begin{eqnarray}
{\cal G} &= & \int d^4x \, \Big [
(\tilde \varphi^0_+)^2 + (\tilde \varphi^a_+)^2
+ \frac{g}{m} \tilde \varphi^0_+ (\tilde e^0 \tilde b^0 + \tilde e^a \tilde b^a)
+ \frac{g}{m} \tilde \varphi^a_+ (
\tilde e^a \tilde b^0
-\tilde e^0 \tilde b^a
- \varepsilon^{abc} \tilde e^b \tilde b^c ) \nonumber \\
&&~~~~~~~ -\frac{g^2}{2m^2}
\Big ( - (2\tilde e^0 \tilde b^0 + \tilde e^a \tilde b^a) \tilde e^b \tilde b^b
+ \varepsilon^{abc} \tilde e^0 \tilde b^b \tilde e^c \tilde b^a
- \varepsilon^{abc} \tilde e^b \tilde b^0 \tilde e^c \tilde b^a
\Big )
\Big ] \, .
\label{25}
\end{eqnarray}
Eq.(\ref{25}) holds in the gauge $\tilde{\varphi}_a^-=0$. It can be
made gauge-invariant by performing the substitution
\begin{eqnarray}
&& \tilde{\varphi}^0_+ \rightarrow
\tilde{\varphi}^0_+ + \frac{g}{2m} \tilde{\varphi}^a_-
\tilde{\varphi}^a_+ \, ,
\qquad
\tilde{\varphi}^a_+ \rightarrow
\tilde{\varphi}^a_+ - \frac{g}{2m} \tilde{\varphi}_-^a
\tilde{\varphi}^0_+ +
\frac{g}{2m} \varepsilon^{abc} \tilde{\varphi}^b_- \tilde{\varphi}^c_+ \, , \nonumber \\
&&
\tilde{e}^0 \rightarrow \tilde{e}^0 + \frac{g}{2m} \tilde{\varphi}^a_- \tilde{e}^a \, ,
\qquad
\tilde{e}^a \rightarrow \tilde{e}^a - \frac{g}{2m} \tilde{\varphi}_-^a
\tilde{e}^0 +
\frac{g}{2m} \varepsilon^{abc} \tilde{\varphi}^b_- \tilde{e}^c \, , \nonumber \\
&&
\tilde{b}^0 \rightarrow \tilde{e}^0 + \frac{g}{2m} \tilde{\varphi}^a_- \tilde{b}^a \, ,
\qquad
\tilde{b}^a \rightarrow
\tilde{b}^a - \frac{g}{2m} \tilde{\varphi}_-^a
\tilde{b}^0+
\frac{g}{2m} \varepsilon^{abc} \tilde{\varphi}^b_- \tilde{b}^c \, .
\label{26}
\end{eqnarray}\
This yields
\begin{eqnarray}
\!\!\!\!\!\!\!\!\!\!\! {\cal G} &= & \int d^4x \, \Big [
\Big ( \tilde \varphi^0_+ + \frac{g}{2m} \tilde \varphi_-^a \tilde \varphi_+^a
+ \frac{g}{2m} ( \tilde e^0 \tilde b^0 + \tilde e^a \tilde b^a) \Big )^2 \nonumber \\
&&  + \Big (\tilde \varphi^a_+ - \frac{g}{2m} \tilde \varphi^0_+ \tilde \varphi^a_-
- \frac{g}{2m} \varepsilon^{abc}  \tilde \varphi_-^b \tilde \varphi_+^c +
\frac{g}{2m} (\tilde e^a \tilde b^0 - \tilde e^0 \tilde b^a -
\varepsilon^{abc} \tilde e^b \tilde b^c) \Big )^2
\Big ] \, .
\label{27}
\end{eqnarray}
A further solution exists
\begin{eqnarray}
{\cal G}_1 & = & \int d^4x \, \Big (
\tilde{\varphi}^0_+ [ ( \varphi - \hat \mu)^* (\varphi-\hat \mu)
- \mu^2]  \nonumber \\
&& ~~~~~~~~~ +
\frac{g}{2m} (\tilde e^a \tilde b^a + \tilde e^0 \tilde b^0)
 [ ( \varphi - \hat \mu)^* (\varphi-\hat \mu)
- \mu^2]  \Big ) \, .
\label{28}
\end{eqnarray}
It can be made gauge-invariant by performing the substitution
eq.(\ref{26}). This yields
\begin{eqnarray}
{\cal G}_1 & = & \int d^4x \, \Big (
\Big ( \tilde{\varphi}^0_+ + \frac{g}{2m} \tilde{\varphi}^a_-
\tilde{\varphi}^a_+ \Big )
[ ( \varphi - \hat \mu)^* (\varphi-\hat \mu)
- \mu^2]  \nonumber \\
&& ~~~~~~~~~ +
\frac{g}{2m} (\tilde e^a \tilde b^a + \tilde e^0 \tilde b^0)
 [ ( \varphi - \hat \mu)^* (\varphi-\hat \mu)
- \mu^2]  \Big ) \, .
\label{29}
\end{eqnarray}
The new effective action becomes
\begin{eqnarray}
A_{eff} = \int d^4x \, \tilde{ L} + \frac{m^2}{2}\alpha {\cal G} +
\frac{m g}{2} \beta {\cal G}_1 \, . \label{30}
\end{eqnarray}
The factors of $m$ in the above equation have been inserted for
dimensional reasons. $\alpha$ and $\beta$ are dimensionless free
parameters.

The quadratic part of $A_{eff}$ is
\begin{eqnarray}
&& \!\!\!\!\!\!\!\!\!\!\!\!\!\!\!\!\!\!\!\!\!\!\!\!\!\!\!\!\!\!\!\!\!\!\!
A_{eff,0} =  \int d^4x \, \Big (
-\frac{1}{4} (\partial_\mu A^a_\nu - \partial_\nu A^a_\mu)^2
+ \frac{\mu^2}{2} (A_{\mu}^{a})^2 +
\mu \varphi^a \partial A^a  \nonumber \\
&&
\!\!\!\!\!\!
+ m \tilde{\varphi}^a_+ \partial A^a
+ \frac{1}{2} \partial_\mu \varphi^a \partial_\mu \varphi^a
+ \frac{\alpha}{2} m^2 (\tilde \varphi^a_+)^2
\nonumber \\
&& 
\!\!\!\!\!\!
+ \frac{1}{2} \partial_\mu \varphi_0 \partial_\mu \varphi_0 -
\frac{M_H^2}{2}\varphi_0^2 + \beta \mu m \varphi_0 \tilde
\varphi_0^+ + \frac{\alpha m^2}{2} (\tilde \varphi_0^+)^2 -\tilde
\varphi_0^+ \square h +\partial_{\mu}\tilde{e}^\alpha
\partial_{\mu}\tilde{b}^\alpha \Big ) \, . \label{31}
\end{eqnarray}
where $\alpha$ ranges over $0,a$.
We get the following
non-vanishing propagators:
\begin{eqnarray}
&& \Delta(A^a_{\mu} A^b_{\nu}) =
i \delta^{ab} \Big ( \frac{1}{-p^2 + \mu^2} T_{\mu\nu}
- \frac{\alpha}{p^2} L_{\mu\nu} \Big ) \, , ~~~
\Delta(A_{\mu}^a \varphi^b) = \delta^{ab}
\frac{i \alpha \mu}{p^4} p_\mu \, , \nonumber \\
&& \Delta(\varphi^a \varphi^b) =
\delta^{ab} \frac{i}{p^4} (p^2 - \alpha \mu^2) \,  \nonumber \\
&& \Delta( A_{\mu}^a \tilde \varphi^b_+) =- \delta^{ab} \frac{p_\mu
}{m p^2}\, , ~~~ \Delta(\tilde{b}^a
\tilde{e}^b)=\frac{i\delta^{ab}}{p^2} \, . \label{32}
\end{eqnarray}
They are the same as those obtained in the Higgs-Kibble model in the
Lorentz covariant $\alpha$-gauge once $\tilde \varphi_a^+$ is
identified with the Nakanishi-Lautrup field. Hence in this
sector the physical fields coincide with the physical fields in the
corresponding sector of the ordinary Higgs-Kibble model, that is
include three components of the massive vector field $A_{\mu}$. 

The non-vanishing propagators in the sector spanned by the fields
$h, \tilde{\varphi}_0^+, \varphi_0$ are given by
\begin{eqnarray}
&& \Delta(hh) = - \frac{i \alpha m^2}{p^4} - \frac{i \beta^2 m^2
\mu^2}{p^4(p^2-M_H^2)} \, , ~~~
\Delta(h \tilde{\varphi}^0_+) = \frac{i}{p^2} \, , \nonumber \\
&& \Delta(h \varphi^0) = -\frac{i \beta \mu m}{p^2(p^2-M_H^2)} \, ,
~~~ \Delta(\varphi_0 \varphi_0) = \frac{i}{p^2 -M_H^2}\,  , ~~~
\Delta(\tilde{b}^0 \tilde{e}^0)=\frac{i}{p^2} \, . \label{33}
\end{eqnarray}

The new terms in eq.(\ref{30}) do not modify the structure of the
nilpotent charge $Q^0$ in eq.(\ref{23}). 
Thus
physical unitarity follows
as in Sect.~\ref{unit} and we conclude that the only physical states
are the three components of the massive gauge field
and the massive scalar Higgs particle. 

Additional divergencies are presented by the tadpole term
\begin{eqnarray}
{\cal T} = \int d^4x \,  \Big [ ( \varphi - \hat \mu)^* (\varphi-\hat \mu)
- \mu^2 \Big ] \, .
\label{38}
\end{eqnarray}
and the similar term for $\tilde{\varphi}_+^0$. To  fulfill all the
symmetries of the theory it must have the form
\begin{eqnarray}
{\cal I}_t = \int d^4x \, \Big ( \frac{g}{2m}
(\tilde e^a \tilde b^a + \tilde e^0 \tilde b^0)
+ \tilde{\varphi}^0_+ + \frac{g}{2m} \tilde{\varphi}^a_-
\tilde{\varphi}^a_+  \Big ) \, .
\label{39}
\end{eqnarray}
So we are finally led to study the following effective action
\begin{eqnarray}
A'_{eff} = \int d^4x \,
  {\cal L} + \frac{m^2}{2}\alpha {\cal G}
+ \frac{m g}{2} \beta {\cal G}_1 + t {\cal T}
+\tilde t {\cal I}_t
\, .
\label{40}
\end{eqnarray}
The invariant in eq.(\ref{38}) allows to adopt the normalization condition on the 1-PI vertex functional
\begin{eqnarray}
\frac{\delta \Gamma}{\delta \varphi_0} = 0
\label{41}
\end{eqnarray}
to all orders in the loop expansion. This is the choice adopted
at tree level in eq.(\ref{1}).

Therefore the physical asymptotic sector includes three
components of the massive vector field and one massive scalar. 

The renormalizability of the theory after inclusion of the terms
(\ref{25},\ref{28}) follows directly from the Feynman rules in the
effective action (\ref{40}) and the propagators (\ref{32},\ref{33})
which lead to the same expression for the divergency index as
before.

\section{Structure of the counterterms}

In this Section we will prove that the UV
divergences of the theory can be removed
only by changing the values of the parameters entering
into the effective action (\ref{40}) (modulo field
redefinitions). Moreover this procedure
does not violate the symmetries of theory, i.e.
the invariance generated by the transformation
in eq.(\ref{21}) and the residual global SU(2) symmetry.

In order to study the structure of the counterterms in the
gauge $\tilde \varphi^a_-=0$ let us introduce the
tree-level vertex functional $\G^{(0)}$, including
apart from the classical action $A'_{eff}$
in eq.(\ref{40}) also the variation of the fields
$\Phi$ under the transformation
(\ref{21}), coupled to the external sources $\Phi^*$
(the antifields \cite{ZJ}).
Then the invariance under the transformation
(\ref{21}) is translated into the following functional
identity
\begin{eqnarray}
{\cal S}(\G) = \int d^4x \sum_\Phi
\frac{\delta \G}{\delta \Phi^*(x)} \frac{\delta \G}{\delta \Phi(x)} = 0
\, .
\label{st.1}
\end{eqnarray}
The vertex functionl $\G$ can be developed in the number
of loops, i.e. $\G = \sum_{j=0}^\infty \hbar^j \G^{(j)}$.

Assuming that a gauge-invariant regularization exists,
the effective action $\hat \G$ including all the counterterms
also fulfills the same functional identity
\begin{eqnarray}
{\cal S}(\hat \G) = 0 \, .
\label{st.2}
\end{eqnarray}
In addition the residual global SU(2) invariance is also
respected. By taking into account power-counting bounds
the most general solution to eq.(\ref{st.2})
which is invariant under the residual global SU(2)
symmetry is obtained from $\G^{(0)}$ upon
redefinition of the free parameters
\begin{eqnarray}
&& g' = Z_g g \, , ~~~ m' = Z_m m \, ,  ~~~ t' = Z_t t \, , ~~~
\tilde t' = Z_{\tilde t} \tilde t \, , ~~~
\alpha' = \frac{Z_\alpha}{Z_g^2} \alpha \, , \nonumber \\
&& \beta' = \frac{Z_\beta}{Z_g Z_m} \beta \, , ~~~
\lambda' = Z_\lambda \lambda
\label{fr.2}
\end{eqnarray}
and by performing a field redefinition
which preserves the residual global SU(2) invariance
and the UV counting
\begin{eqnarray}
&& \tilde e' = Z_1 \tilde e \, , ~~~~  \tilde b' = Z_m \tilde b \, , ~~~~  A^{a'}_\mu = Z_2 A^a_\mu \, ,
~~~~ h' = Z_m Z_3 h \, , \nonumber \\
&& \varphi^{0'} = z_1 \varphi^0 \, , ~~~
      \varphi^{a'} = z_1 \varphi^a \, , \nonumber \\
&& \tilde \varphi^{a'}_+ =
Z_4 \tilde \varphi^a_+ +
Z_5 \partial A^a + Z_6 \frac{1}{m} \partial_\mu h A^{a\mu}
+ Z_7 (\tilde e^0 \tilde b^a - \tilde e^a \tilde b^0 -
\varepsilon^{abc} \tilde e^b \tilde b^c ) \, , \nonumber \\
&& \tilde \varphi^{0'}_+ =
Z_8 \tilde \varphi^0_+ + Z_9 \frac{1}{m} \square h +
Z_{10} \frac{1}{m^2} \partial_\mu h \partial^\mu h + Z_{11} A^2
+Z_{12} (\tilde e^0 \tilde b^0 + \tilde e^a \tilde b^a) \nonumber \\
&& ~~~~~~~+ z_2 \Big [ (\varphi_0 + \mu) ^2 + \varphi_a^2 - \mu^2 \Big ]
\, .
\label{fr.3}
\end{eqnarray}
In the above equation we have kept the notation of capital $Z$'s for
those field redefinition constants which are common with the
massless YM case \cite{QS}. Moreover we have introduced small $z$'s
for the new field redefinitions. One should notice that the residual
global SU(2) symmetry imposes several non-trivial constraints: it
forbids a term proportional to the gradient of $\varphi^a$ in the
redefinition of the gauge field $A^a_\mu$ (which on the contrary is
present in the standard 't Hooft gauge since the latter
breaks explicitly the residual global SU(2) symmetry) and excludes
the appearance of terms proportional to $\varphi^a$ and $\varphi^0
\varphi^a$ in the redefinition of $\tilde{\varphi}^a_+$, which would
be
 otherwise allowed by the power-counting.
Moreover it selects the invariant
$ ((\varphi_0+\mu)^2 + \varphi_a^2 - \mu^2) $
as the unique combination which can enter
the field redefinition of $\tilde \varphi^0_+$.

The field redefinition in eq.(\ref{fr.3}) must be implemented
without violating the functional identity  (\ref{st.2}).
A convenient way is to make use of canonical transformations
as in \cite{QS}. For that purpose
we rewrite the functional
identity (\ref{st.2}) by means of the
following bracket \cite{Troost:1989cu}
\begin{eqnarray}
(X,Y) = \int d^4x \sum_\Phi (-1)^{\epsilon(\Phi)\epsilon(X)}
\Big (
\frac{\delta X}{\delta \Phi} \frac{\delta Y}{\delta \Phi^*}
- (-1)^{\epsilon(X)+1} \frac{\delta X}{\delta \Phi^*}\frac{\delta Y}{\delta \Phi} \Big )
\label{bracket}
\end{eqnarray}
where $\epsilon$ denotes the statistics
($1$ for fermions, $0$ for bosons). 
In terms of the bracket (\ref{bracket}) one has
\begin{eqnarray}
{\cal S}(\G) = \frac{1}{2} (\G,\G) = 0 \, .
\label{bracket.fi}
\end{eqnarray}
Under eq.(\ref{bracket}) the fields and the antifields are paired via the
fundamental brackets
\begin{eqnarray}
(\Phi_i,  \Phi^*_j ) =(-1)^{\epsilon(\Phi_j)} \delta_{ij} \, .
\label{bracket.fund}
\end{eqnarray}
The conventions on the antifields differs from the one
of \cite{Troost:1989cu} by the redefinition $\Phi^* \rightarrow
(-1)^{\epsilon(\Phi)} \Phi^*$, whence the sign factor in the r.h.s.
of the above equation.

A redefinition of the fields and the antifields respecting eq.(\ref{bracket.fund})
 preserves the bracket between any two functionals $X,Y$ and hence also the functional identity (\ref{st.2}). 
Such a redefinition
is called a canonical transformation (w.r.t. the bracket (\ref{bracket})).

The easiest way to work out the appropriate
canonical transformation is to make use
of the finite canonical transformation
generated by the functional
 $G = \int d^4x \sum_{\Phi'}
(-1)^{\epsilon(\Phi')} \Phi^{*'} \Phi'(\Phi)$ ~\cite{Troost:1989cu}.
The field transformations fix the dependence of $G$ on
the new antifields, while the antifield redefinitions are
obtained
 by solving the equations
\begin{eqnarray}
\Phi^* = (-1)^{\epsilon(\Phi)} \frac{\delta G}{\delta \Phi} \, , ~~~
\Phi' = (-1)^{\epsilon(\Phi')} \frac{\delta G}{\delta \Phi^{*'}} \, .
\label{sc.fin}
\end{eqnarray}
By explicit computation one finds
\begin{eqnarray}
&&
{\tilde e}^{0*} = Z_1 {\tilde e}^{0*'}
                            + Z_7 {\tilde \varphi}^{a*'}_+ {\tilde b}^a
                            + Z_{12} {\tilde \varphi}^{0*'}_+ {\tilde b}^0 \, , \nonumber \\
&&
 {\tilde e}^{a*} = Z_1 {\tilde e}^{a*'} -
                 Z_7   {\tilde \varphi}^{a*'}_+ {\tilde b}^0 +
                 Z_7 \varepsilon^{abc}  {\tilde \varphi}^{b*'}_+ {\tilde b}^c
                 + Z_{12} {\tilde \varphi}^{0*'}_+ {\tilde b}^a \, , \nonumber \\
&&
{\tilde A}^{a*}_\mu = Z_2 A^{a*'}_\mu
                 - Z_5 \partial^\mu {\tilde \varphi}^{a*'}_+
                 + \frac{Z_6}{m} \partial_\mu h
                 {\tilde \varphi}^{a*'}_+
                 + 2 Z_{11} {\tilde \varphi}^{0*'}_+ A^a_\mu \, , \nonumber \\
&& \varphi^{0*} = z_1 \varphi^{0*'}
+2 z_2 (\varphi_0 + \mu) \varphi^{0*'}
\, , ~~~~
\varphi^{a*} = z_1 \varphi^{a*'}
+ 2 z_2 \varphi^a \varphi^{0*'}
\, , \nonumber \\
&&
 {\tilde \varphi}^{a*}_+ = Z_4  {\tilde \varphi}^{a*'}_+ \, , ~~~~  {\tilde \varphi}^{0*}_+ = Z_8  {\tilde \varphi}^{0*'}_+ \, , \nonumber \\
&&
{\tilde b}^{a*} = - Z_7  {\tilde \varphi}^{a*'}_+ \tilde e^0
                  + Z_7 \varepsilon^{abc}   {\tilde \varphi}^{b*'}_+ \tilde e^c
                  - Z_{12}  {\tilde \varphi}^{0*'}_+ \tilde e^a + Z_m {\tilde b}^{a*'}  \, , \nonumber \\
&&
 {\tilde b}^{0*} = Z_7 {\tilde \varphi}^{a*'}_+ \tilde e^a -
                  Z_{12}  {\tilde \varphi}^{0*'}_+ \tilde e^0 + Z_m {\tilde b}^{0*'} \, , \nonumber \\
&&
h^* = -\frac{Z_6}{m} {\tilde \varphi}^{a*'}_+ \partial A^a
- \frac{Z_6}{m}  \partial_\mu {\tilde \varphi}^{a*'}_+ A^{a\mu}
+ \frac{Z_9}{m} \square {\tilde \varphi}^{0*'}_+ \nonumber \\
&&
~~~~~~ - 2\frac{Z_{10}}{m^2} \square h {\tilde \varphi}^{0*'}_+
-2\frac{Z_{10}}{m^2} \partial^\mu  {\tilde \varphi}^{0*'}_+ \partial_\mu h
+ Z_m Z_3 h^{*'} \, .
\label{af.1}
\end{eqnarray}
Consequently the functional
\begin{eqnarray}
&&
\!\!\!\!\!\!\!\!\!\!\!\!\!\!\!\!\!\!\!\!
\hat \G[ g',m',t',{\tilde t}', \alpha',\beta',\lambda'; \Phi', \Phi^{*'}] =
 \G^{(0)}[Z_g g, Z_m m,
 Z_t t,  Z_{{\tilde t}'} \tilde{t}, Z_\alpha/Z_m^2 \alpha,
 Z_\beta /Z_g Z_m \beta, Z_\lambda \lambda
 ; \nonumber \\
 && \qquad \qquad \qquad \qquad \qquad \qquad
 \Phi(\Phi'),\Phi^{*}(\Phi',\Phi^{*'})]
\label{sc.5}
\end{eqnarray}
is the most general solution to eq.(\ref{st.2})
compatible with power-counting bounds. One can verify it by
explicit calculations.

It finally remains to be shown that the UV divergences can be recursively
reabsorbed by a change in the parameters  $Z_g,Z_m,Z_t,Z_{{\tilde
t}'},Z_\alpha,Z_\beta,Z_\lambda$ and  field renormalization
constants $Z_j$, $j=1,\dots,12$, $z_j$, $j=1,2$, order by order in
the loop expansion. This technical proof is left to Appendix A.

\section{Comparison with the usual formulation.}

We proved above that the renormalized Higgs-Kibble model is
described by the gauge invariant Lagrangian and generates the
scattering matrix which is unitary in the space including three
components of the massive Yang-Mills field and one massive scalar
particle.

Now we show that in the framework of perturbation theory the
scattering matrix obtained above may be transformed to the ususal
renormalizable gauges, or to nonrenormalizable unitary gauge. The
comparison of our formulation with the standard one exactly
coincides with the corresponding procedure for the Yang-Mills theory
\cite{QS}. For that reason we shall not repeat the proof. As in the
Yang-Mills case our scattering matrix may be presented as the path
integral
\begin{equation} S=\int
\exp\{i\int[L_{g.i.}(x)+\lambda^a\varphi^a_-]d^4x\}\det(M^{ab}) d\mu \, ,
 \label{cmp.1}
\end{equation}
where $L_{g.i.}$ denotes the gauge invariant Lagrangian entering the
effective action (\ref{40}), and
\begin{equation}
\det(M^{ab})^{-1}|_{\varphi^a_-=0}=\int\delta((\varphi^{\Omega}_-)^a)
d\Omega \label{cmp.2}
\end{equation}
and the vacuum  boundary conditions may be adopted for the
auxilliary fields $\varphi_{\pm}, e, b$. Multiplying the integral
(\ref{cmp.1}) by "1"
\begin{equation}
1=\Delta_L\int\delta(\partial_{\mu}A_{\mu}^\Omega) d\Omega
\label{cmp.3}
\end{equation}
and changing the variables $\Phi^\Omega\rightarrow \Phi$ we arrive
to the expression for the scattering matrix in the gauge
$\partial_{\mu}A_{\mu}=0$:
\begin{equation}
S=\int
\exp\{i\int[L_{g.i.}(x)+\lambda^a(x)\partial_{\mu}A_{\mu}^a(x)+\partial_{\mu}\bar{c}^a(D_{\mu}c)^a]d^4x\}d\mu
\label{cmp.4}
\end{equation}

As the vacuum boundary conditions were adopted for the
fields $\varphi_{\pm}, e, b$,  we can integrate out all these
fields and obtain the standard expression for the scattering matrix
of the Higgs-Kibble model in the gauge $\partial_{\mu}A_{\mu}=0$.
In the same way one can consider (again in the framework of
perturbation theory) other renormalizable gauges.

Finally we mention that the independence on the choice of the gauge
holds also for expectation values of other gauge invariant
operators.

\section{Discussion}

In this paper we showed that the Higgs-Kibble model may be
formulated in the close analogy with the pure Yang-Mills  theory.
The corresponding theory is renormalizable and ambiguity
free. Hence we conclude that the appearance of the ambiguity in the
standard procedure is the artefact of the quantization procedure. Of
course the final answer to the question of the physical importance
of the Gribov ambiguity may be given only beyond perturbation
theory.

\vskip 0.5 truecm
{\bf Acknowledgements.}

\vskip 0.2 truecm
One of us (A.Q.) gratefully acknowledges
partial financial support from University of Milano.
 The work of A.A.S. was partially
supported by Russian
 Basic Research Foundation under grant
 09-01-12150-{\cyr ofi\_m} and RAS program "Nonlinear dynamics".

\appendix
\section{Recursive Removal of the UV Divergences}

In this Appendix we show that the UV divergences
of the model can be reabsorbed order by order in the
loop expansion by a suitable choice of the
parameters $Z_g,Z_m,Z_t,Z_\alpha,Z_\beta,Z_\lambda$ and  field renormalization constants
$Z_j$, $j=1,\dots,12$, $z_j$, $j=1,2$.

The proof closely parallels the one already presented
in \cite{QS} for the massless Yang-Mills theory and thus
we will only sketch the main points here.

Suppose that the subtraction of the divergences
has been performed up to order $n-1$
in the loop expansion while preserving
the  residual global
$SU(2)$ invariance and eq.(\ref{st.1}).
Then at order $n$ eq.(\ref{st.1}) gives
\begin{eqnarray}
&& {\cal S}_0(\G^{(n)}) \equiv
\int d^4x \, \sum_\Phi \Big ( \frac{\delta \G^{(0)}}{\delta \Phi^*(x)}
\frac{\delta}{\delta \Phi(x)} +
\frac{\delta \G^{(0)}}{\delta \Phi(x)}
\frac{\delta}{\delta  \Phi^*(x)}
\Big ) \G^{(n)} =  \nonumber \\
&& ~~~~~~~~~~~
- \sum_{j=1}^{n-1}
\int d^4x \, \sum_\Phi
\frac{\delta \G^{(n-j)}}{\delta \Phi^*(x)}
\frac{\delta \G^{(j)}}{\delta \Phi(x)}
\label{comm.3}
\end{eqnarray}
The second line of the above equation is finite since
it contains only lower order terms which have
already been subtracted. Hence
one gets the following equation for the divergent
part $\G^{(n)}_{div}$ at order $n$
\begin{eqnarray}
{\cal S}_0 (\G^{(n)}_{div}) = 0 \, .
\label{comm.4}
\end{eqnarray}
The operator ${\cal S}_0$ 
is defined by the first line of eq.(\ref{comm.3}).
Since the antifield $\Phi^*$ is coupled in $\G^{(0)}$ 
to the transformation of the field $\Phi$ in eq.(\ref{21}),
the ${\cal S}_0$-variation of $\Phi$ coincides with
$\delta \Phi$ in eq.(\ref{21}):
\begin{eqnarray}
{\cal S}_0 \Phi = \frac{\delta \G^{(0)}}{\delta \Phi^*} =
\delta\Phi \, .
\label{comm.5}
\end{eqnarray}
From eq.(\ref{comm.3}) one also sees that ${\cal S}_0$  acts on the antifield
$\Phi^*$ by mapping it into the classical e.o.m.
of $\Phi$, namely
\begin{eqnarray}
{\cal S}_0 \Phi^* = \frac{\delta \G^{(0)}}{\delta \Phi} \, .
\label{comm.6}
\end{eqnarray}
${\cal S}_0$ is nilpotent, as a consequence of
the nilpotency of $\delta$ and the validity of
the functional identity eq.(\ref{st.1}) for $\G^{(0)}$.

\medskip
The most general solution to
eq.(\ref{comm.4}) can be written as
\begin{eqnarray}
\G^{(n)}_{div} = A + {\cal S}_0 B
\label{comm.7}
\end{eqnarray}
where $A$ cannot be presented in the form ${\cal S}_0 C$, with $C$ a
local functional.
$\G^{(n)}$ is invariant under the global residual
$SU(2)$ symmetry preserved by the regularization.
 Since $\Gamma^{(n)}$ is a Lorentz
invariant functional, the functionals $A$ and $B$ also possess this
invariance. 

There is a general strategy for obtaining the most
general solution of the A-type. This relies on
the evaluation of the
cohomology
$H_{{\cal F}}({\cal S}_0)$ \cite{HCohom}  of the nilpotent operator ${\cal S}_0$
in the space ${\cal F}$ of Lorentz- and global $SU(2)$-invariant local functionals
with dimension bounded by the power-counting.
$H_{\cal F}({\cal S}_0)$ is defined as the quotient of the latter functional space w.r.t. to the equivalence
relation
\begin{eqnarray}
X \sim Y \Leftrightarrow {X}-{Y}={\cal S}_0(C)
\label{equiv}
\end{eqnarray}
for some  Lorentz- and global $SU(2)$-invariant local functional $C$.

%%%%%%%%%%%%%%%%%%%%%%%%%%%%%%%

The first step in the computation of ${\cal H}_{\cal F}({\cal S}_0)$ is the identification of the so-called doublet variables.
A pair of variables $u,v$ such that
${\cal S}_0 u = v \, , {\cal S}_0 v =0$
is called a  ${\cal S}_0$-doublet.
Their importance stems from the fact that 
the dependence on $u,v$
can only happen via the term ${\cal S}_0B$
in eq.(\ref{comm.7}), as a consequence of 
a general
theorem valid for nilpotent differentials \cite{Barnich:2000zw, Quadri:2002nh}.
It is easy to see that the pairs $(h, -\tilde b^0)$,
$(\tilde e^a, \frac{g}{2m} (\tilde e^a \tilde b^0
- \tilde e^0 \tilde b^a - \varepsilon^{abc} \tilde e^b \tilde b^c) +
\tilde{\varphi}^a_+)$,
$(\tilde e^0, \frac{g}{2m} (\tilde e^a \tilde b^a 
+ \tilde e^0 \tilde b^0) + \tilde{\varphi}^0_+)$
satisfy the definition of ${\cal S}_0$-doublets.
Moreover the doublet partners 
of $\tilde e^0, \tilde e^a$ are in one-to-one
correspondence with $\tilde{\varphi}^0_+,
\tilde{\varphi}^a_+$ (via an invertible
field redefinition).
Hence we conclude that  the dependence of the most general solution
to eq.(\ref{comm.4}) on $h,\tilde e^a, \tilde e^0$ and
$\tilde b^0, \tilde{\varphi}^a_+,
\tilde{\varphi}^0_+$ is confined to the term ${\cal S}_0B$.
We remark that a similar result also holds in the massless
Yang-Mills theory  in the
ambiguity-free gauge \cite{QS}.

Since the dependence on the doublets
is confined to the ${\cal S}_0 B$-sector,
we can now consider
 the restriction of ${\cal S}_0$
to the subspace without doublets
(and their antifields).

In this latter subspace
the action of ${\cal S}_0$ is the same as the one of the standard gauge BRST transformation of the Higgs model for the
gauge group $SU(2)$, once one identifies the $SU(2)$ BRST ghosts
with $\frac{1}{m} \tilde b_a$,
as was done in eq.(\ref{21}).
%%%%%%%%%%%%%%%%%%%%%%%%%%%%%%%
%%%%%%%%%%%%%%%%%%%%%%%%%%

By taking into account power-counting bounds, the class of $A$-type solutions
is exhausted by the gauge-invariant
polynomials in the gauge field $A^a_\mu$
and the Higgs doublet $\varphi$ of
dimension less or equal than four \cite{HCohom}, i.e.
\begin{eqnarray}
&& \G^{(n)}_{div} = \int d^4x \,
\Big ( a_1^{(n)} F^a_{\mu\nu}F^a_{\mu\nu}
+ a_2^{(n)} (D_\mu \varphi)^* (D_\mu \varphi) \nonumber \\
&& ~~~~~~~~~~~~~~~~
+ a_3^{(n)} (\varphi^*\varphi - \mu^2)^2
+ a_4^{(n)} (\varphi^*\varphi - \mu^2) \Big ) \, .
\label{higgs.ord.n}
\end{eqnarray}
The divergent coefficients $a_j^{(n)}$ can be reabsorbed
by a redefinition of $Z_g$, $z_1$, $Z_\lambda$
and $Z_t$ respectively.

The enumeration of the solutions of the type
${\cal S}_0B$ goes as follows.
There are two classes of this kind of solutions:
those which do not depend on the antifields
and those which depend on the antifields.
The first ones are obtained in terms of the following operator insertions
\begin{eqnarray}
&&
\!\!\!\!\!\!\!\!\!\!\!\!\!\!\!\!\!\!\!\!\!\!\!
{\cal J}_m = \delta Z^{(n)}_m \frac{\partial}{\partial Z_m} \hat \G \, , ~~ {\cal J}_\alpha = \delta Z^{(n)}_\alpha \frac{\partial}{\partial Z_\alpha} \hat \G \, , ~~
{\cal J}_{\beta} = \delta Z^{(n)}_\beta \frac{\partial}{\partial Z_\beta} \hat \G \,, ~~  {\cal J}_{\tilde t} = \delta Z^{(n)}_{\tilde t} \frac{\partial}{\partial Z_{\tilde t}} \hat \G \, ,
\label{b.solutions}
\end{eqnarray}
where $\delta Z^{(n)}_m$, $\delta Z^{(n)}_\alpha$,
$ \delta Z^{(n)}_\beta$,
$\delta Z^{(n)}_{\tilde t}$ are divergent coefficients of order $n$.
They can be reabsorbed by a redefinition
of $Z_m,Z_\alpha, Z_\beta,  Z_{\tilde t}$ respectively.

\medskip
The invariants of the ${\cal S}_0 B$-type involving the antifields
are of the form (no sum over $\Phi$)
\begin{eqnarray}
w^{(n)}_j \int d^4x \, {\cal S}_0 ( \Phi^* F_j(\Phi)) =
w^{(n)}_j \int d^4x \Big ( F_j(\Phi) \frac{\delta \G^{(0)}}{\delta \Phi}
- \Phi^* \frac{\delta F_j}{\delta \Phi} {\cal S}_0 \Phi \Big )
\label{comm.8}
\end{eqnarray}
where again the divergent coefficients $w_j^{(n)}$ are
of order $\hbar^n$.The possible $F_j$'s, $j=1,\dots,14$
in eq.(\ref{comm.8}) are constrained
by the rigid symmetries of the theory,
quantum numbers and power-counting
and have the same structure as the
corresponding terms in eq.(\ref{fr.3}).
They can be reabsorbed by a $n$-th order redefinition of the
field renormalization constants $Z_j$, $j=1,\dots,12$
and $z_j$, $j=1,2$,  as it can be seen
from the first term in the r.h.s. of eq.(\ref{comm.8}).
The corresponding redefinition of the antifields 
in the second term of eq.(\ref{comm.8}) is automatically taken
into account by the associated antifield
redefinition in eq.(\ref{af.1}).

Once the divergences have been symmetrically removed up to order
$n$, the procedure can be iterated at order $n+1$.
This completes the proof that indeed renormalization
does not spoil eq.(\ref{st.1}) and the residual global
SU(2) invariance. Therefore we conclude that the renormalized theory
is unitary in the subspace including the
transverse degree of freedom of the massive gauge field
and the physical scalar Higgs mode.

\end{document}